# X-ray spectroscopy technique for the pile-up region

Gaurav Sharma[1], Deepak Swami[2], Basu Kumar[3], N. K. Puri[1] and T. Nandi[2]

[1]*Delhi Technological University, Bawana Road, Delhi 110042, India*

[2]*Inter University Accelerator Centre, Aruna Asaf Ali Marg, New Delhi 110067, India*

[3]*Jai Prakash Vishvidyalaya, Chapra, Bihar 841301, India*

We report a pile-up rejection technique based on X-ray absorption concept of Beer-Lambert's law for measuring true events in the pile-up region. We have detected a $10^4$ times weaker peak in the pile-up region. This technique also enables one to resolve the weak peaks adjacent to an intense peak provided the later lies in the lower energy side, and the peaks are at least theoretically resolvable by the detector used. We have resolved such peaks by reducing the intensity ratios in our experiment. The technique allows us to obtain the actual intensities of the observed peaks to have been measured without any attenuator. The possible applications of this technique can be to study the physics of two electron one-photon transition as well as the properties of projectile-like or target-like ions.

At low bombarding energies, ion-solid/atom collisions lead to a formation of two K-shell vacancies in either target atoms or projectile ions[1]. The two vacancies are mostly filled sequentially by emitting a $K_\alpha$ or $K_\beta$ hypersatellite line and another $K_\alpha$ or $K_\beta$ line[2]. However, there is a small probability that the two vacancies can be filled simultaneously. Accordingly it has been observed that two vacancies can be filled simultaneously by two electrons leading to an emission of a single photon having energies nearly twice that of the diagram lines $K_\alpha$ or $K_\beta$[2–5]. Hence, the signature of two electron one-photon transitions can only be obtained from the spectral lines appearing in the region affected by pile-up phenomenon in the form of $K_{\alpha\alpha}$ or $K_{\alpha\beta}$ lines. Several pile-up rejection techniques including pulse shape discrimination[6], pile-up rejection circuits, reduction of the counting rates, etc [7–9] are known. One important point is to note that the relative yield for $K_{\alpha\alpha}/K_\alpha$ is in the order of $10^{-4}$ in Ar atom[10]. This ratio improves towards lighter atom side and becomes worse on the heavier side[3]. Thereby, the former two techniques capable of handling high counting rates can only be used for these experiments. Nevertheless, these techniques cannot reject the pile-up contribution fully because pile-up rate goes as $n^2\tau$, where n is true count rate and $\tau$ is system dead time and $\tau$ cannot be reduced beyond a certain limit. Most appropriate will be not to restrict counting rate in the experiment but to limit the event rate entering the detector judiciously. Further, in a case of two or more closely spaced spectral lines are present in the spectrum, and one of the line intensity is much stronger than the others, leads to a severe problem in resolving the lines even though detector is capable of resolving them. Such a situation appears in an experiment when intense x-rays from the projectile ions and weak X-ray lines from the projectile-like ions coexist in the spectrum[11,12]. We report a method that can solve both the problems in one go.

The maximum resolution can be achieved provided overlapping of the peaks is made the least. This overlapping can be minimized by either increasing the spectral separation between the peaks or by improving the detector resolution. The former is normally beyond one's capacity. However, the lifetime of the upper levels of the lines may very differently. One group falls in the short time scale and the other on the long time scale. For example, metastable states are long-lived while allowed ones are short lived. One can avoid the short-lived lines by recording the spectra at a delayed condition. This delay is obtained by shifting the source away from the detector in the lifetime measurement using the time of flight method[13]. Another method for reducing some of the peaks is done by using absorption spectroscopy e.g. Doppler-tuned spectroscopy[14]. The first method can be only applied if the lifetime of the excited levels corresponding to the different lines differs significantly, while in the second method, a specific absorber is required for a particular x-ray line.

Both the pile-up problem and unresolved closely spaced strong and weak spectral lines can be solved by attenuating the x-ray photons before they approach the detector. Exactly, by this principle, we have developed a technique that handles high counting rate in the experiment but low counting rate in the detector. In this letter, we report the novel features of x-ray spectroscopy around the pile-up region by utilizing Beer-Lambert's law and strong variation of attenuation constant with photon energy. The practical outcome of such idea is realized in the present experiment.

The experiment was done at IUAC, New Delhi, India using the 15UD palletron accelerator. An ion beam of 110 MeV $Ti^{8+}$ beam was passed through C target of thickness $80\mu g/cm^2$ to create different excited levels in highly charged ions. The energy of the Ti beam was selected above the Coulomb barrier energy so that the X-ray peaks of projectile like ions could also be produced along with different x-ray lines including two electron one-photon transitions. The self-supported C foils prepared using the electron gun evaporation technique was annealed at 900 K in the $N_2$ environment to obtain better uniformity and mechanical stability. The C target on Al holder was mounted on a lifetime setup having a translational accuracy better than 1μm that could move the target to and fro up to 50mm along the beam axis from the detector window. Two low energy Germanium X-ray detectors were placed outside the vacuum at an angle of ± 90º to the beam direction. The X-rays were collected from beam-foil source through a 6μm thick Myler window. Two collimators of diameter 3 and 6 mm at a distance of 50mm were placed in front of the detector. A Faraday cup with electron suppressor was used to normalize the X-ray spectra during lifetime measurements. Aluminum absorbers of different thicknesses were placed in front of the X-ray detectors so that the setup remained undisturbed when the absorber thickness was changed. The vacuum in the chamber was maintained in the range of $10^{-7}$ torr throughout the experiment.

As we mention above that we cannot reduce the counting rate because our main objective to observe the weak lines falling around the pile-up region. Accordingly the recorded spectrum as shown in Fig. 1(a) displays only a peak and a hump like structure at its pile-up region. Now let us see the effect of absorber foils on the spectrum. We have made use of commercially available Al foils of 10.02μm thickness as the x-ray attenuator. Different numbers of absorber foil are put in front of the X-ray detector, and some of the recorded spectra are shown in Fig.1. A clear difference is seen while comparing the spectra with and without the absorbers. We keep on increasing the absorber foil thickness and finally succeeded in getting the peaks detected in the pile-up region as shown in the inset of Fig. 1(c) - 1(e) during eight or more numbers of layers of the Al absorber used. Further, we have measured the intensity decay as a function of distance upstream the beam direction. The delayed spectra feature the same structure as obtained with no delay. It clearly indicates that the spectral features are originating neither from the target contamination nor radiative electron capture processes. The X-ray lines detected are emanating from projectile and projectile like ions, but as the identification of the lines are out of our concern for this letter thus the spectral identification has not been discussed here.

The intensities of the individual peaks are substantially reduced with eight or more numbers of Al layers to obtain all the peaks involved in the spectra well resolved. At this condition we find the intensities, centroid and full width at half maximum (FWHM) of the peaks are consistent with different numbers of Al layers. In the next step, to evaluate the original intensities without any absorber, we have corrected the attenuated intensities for different number of Al layers. One set of such data for 10 absorber layers is shown in Table I. Note that all the peaks lie in the range of 4-10 keV and the efficiency of the detector remains constant in this energy range[15].

As we know the hump like structure in Fig. 1(a) spectrum comes at about the pile-up position of the only one intense peak observed. We can see from Table I that the intensity of this intense peak is reduced by 500 times. Such a low intensity cannot produce any pile-up effect at all. Therefore, the peaks observed in the pile-up region by applying eight or more numbers of absorbers are free from any pile-up effects as shown in Fig 1(c)-1(e). Hence, these spectra represent the real transitions taking place in the ion-solid collisions and not at all any contribution from the pile-up. Whereas the hump-like structure in the Fig. 1(a) is a result of the contributions from both the pile-up due of intense peaks present at lower channels and two atomic transitions. The two peaks are present in that region are possibly from the phenomenon of two electron one photon. Interestingly, the pile-up channel number (~2588ch) corresponding the first peak (~1294ch) falls in between the two peaks appearing at the channel numbers 2483 and 2748. Thus, it reveals that the two peaks measured at the pile-up region are genuine. Further, these two atomic transitions have tiny intensities as can be noticed in Table I. Hence the hump like structure observed in Fig. 1(a) is mostly due to the pile-up contribution of the sole peak at lower channels.

So far we have discussed the high energy region of the spectrum to eliminate the pile-up effects. Now let us discuss how the peaks in the low energy region are deconvoluted from the spectrum shown in Fig. 1(a) using the same X-ray absorber techniques. Theoretically a detector can efficiently resolve two peaks if they are separated by the sum of half width half maxima of the two peaks. In spite of this condition is met by a detector, one experimental limitation can often be encountered if even one of the peaks is much stronger than the others. Then the low intense peaks tend to be veiled in the broad base of the intense peak as shown in a simulated spectrum containing only two peaks having an intensity ratio of 10:1 (see Fig. 2(a)) and consequently they cannot be resolved. Interestingly, this problem can be circumvented by decreasing the relative intensity ratios substantially; a few representative cases displayed in Fig 2(b & c). Unfortunately, we do not have any control over the peak intensities; consequently it is difficult to overcome this problem practically using the same detector. Thus, many low intense peaks remain unidentified even though they are within the resolving capacity of the detector. Figure 2 explains the complete scenario to have the resolution better in Fig.1(c) - 1(e). For resolving such peaks, our technique of X-ray absorption works well provided (a) the intense peak lies at lower energies and (b) they are theoretically resolvable by the detector.

Our technique is solely dependent on the Beer-Lambert's law of absorption of X-ray intensities. Further, the absorption coefficient varies as $E^{-3}$, where E is the energy of incident photons. This dependence of absorption on E gives us an excellent tool to handle the intensities of the different energies differently. This method indicates that the higher energy peaks will be reduced less in comparison to lower energy peaks. If the intense peak has lower energy than its unresolved component, then the ratio of the peaks can be made comparable by selecting the appropriate absorber and its thickness. Such a situation is exactly achieved in the present experiment and three peaks at the lower channel side have clearly been resolved in the spectra shown in Fig. 1(c)-(e).

Using the intensities, centroid and FWHM given in Table I we have tried to simulate the spectrum given in Fig. 1(a). The spectrum so obtained is shown in Fig.3 with an inset to showing last two peaks. The initial structure of Fig. 1(a) and Fig. 3 is quite similar. However, clear dissimilarity can be noticed on comparing pile-up region of the two figures Fig. 3 and Fig. 1 (a). The absence of hump like structure on the high channel number side in Fig. 1(a) is fully missing from Fig. 3 as we have not put in the pile-up contribution in the Fig. 3 spectrum.

To conclude, we have developed a pile-up rejection technique based on X-ray absorption for measuring true events in the pile-up region. This technique also enables us to resolve the weak peaks adjacent to an intense peak provided (a) the intense peak lies in the lower energy side and (b) the peaks are theoretically resolvable by the detector. We have resolved such peaks in our experiment and also detected a $10^4$ times weaker peak in the pile-up region. The technique so developed will be utilized well to investigate the physics of the phenomenon concerning (i) two electron one photon transition as well as the properties of highly charged ions produced by the nuclear reactions in the form of projectile-like, target-like ions, or both[11,16].

**Acknowledgements**

Authors are thankful to the M.Sc students Mr. Sarvesh Sharma and Mr. Udai Singh for their kind help throughout the experiment. Mr. Gaurav Sharma acknowledges UGC for providing the financial support of his research work. We acknowledge DST for funding the project No SP/S2/LOP-0006/2011.

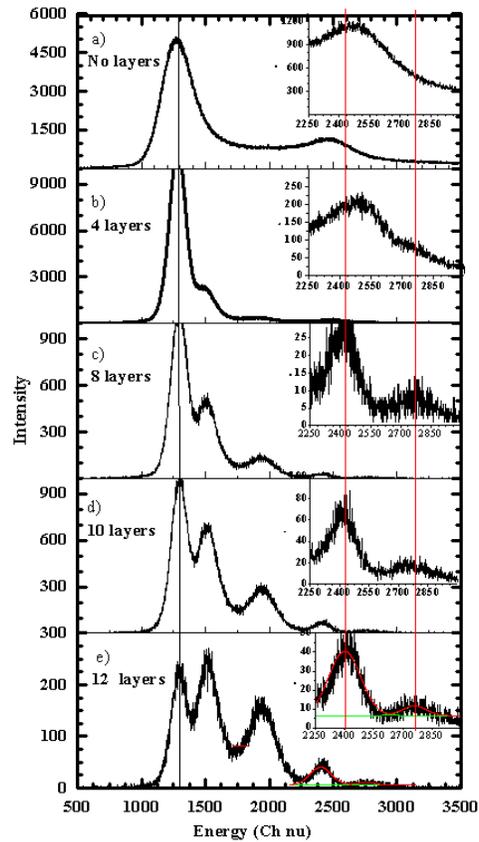

**FIG. 1.** The X-ray spectra with the different absorber foil thickness: (a)-(e) represent the spectra taken with different numbers of Al absorber foils. The measured spectra show improvement on the resolution as a function of absorber foil thickness.

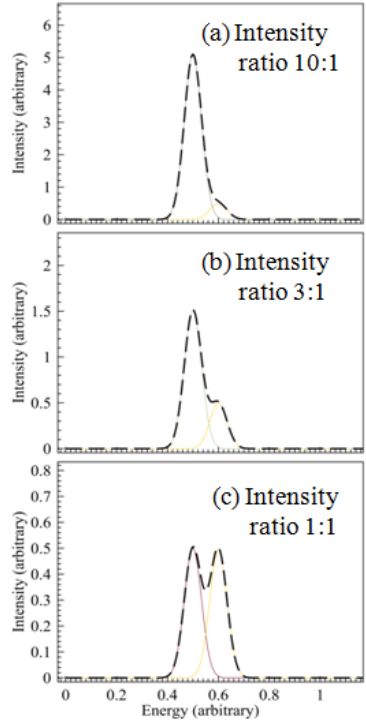

**FIG. 2.** Resolution vs relative intensity plots: Two peaks that are supposed to be resolved by a detector, but are not practically resolved because one is much more intense than the other. How the relative intensity plays the role on the resolution is shown for different intensity ratios keeping the centroid positions and FWHM constant: (a) for 10:1, (b) for 3:1 and (c) for 1:1. The figure thus shows that as the intensity ratio decreases the resolution increases.

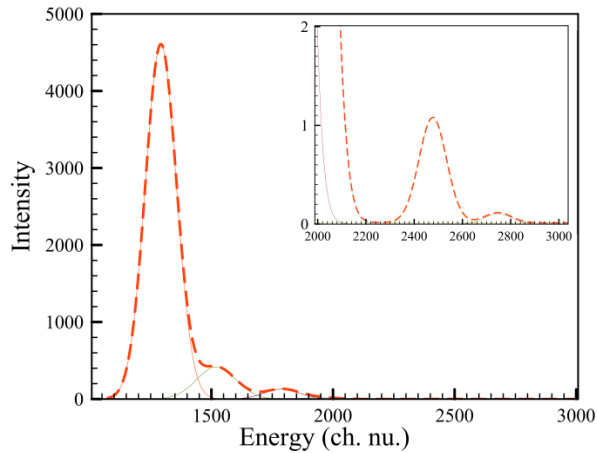

**FIG. 3.** The simulated X-ray spectrum: The spectrum is simulated by using the actual X-ray peak intensities as given in Table I. Since Table I does not include pile-up contribution; no indication of pile-up effect is seen at all in the pile-up region.

**TABLE I.** Attenuation corrected intensities: The X-ray intensities of the measured peaks with ten layers of Al absorbers (100μm) are corrected to have the actual intensities. The figures in the 5$^{th}$ column indicate that pile-up effect is totally nullified and relative intensities as low as 1:2x10$^4$ can well be measured.

| Channel number (keV) | Attenuated Intensity | Transmission % | Corrected Intensity | Intensity Ratio |
|---|---|---|---|---|
| 1293.7 (4.73) | 144957 | 0.20 | 72624098 | 1.00E+00 |
| 1508.7 (5.5) | 145425 | 1.72 | 8441733 | 1.16E-01 |
| 2023.7 (7.36) | 191392 | 8.13 | 2352843 | 3.24E-02 |
| 2482.5 (9.02) | 77210 | 17.15 | 450235 | 6.20E-03 |
| 2748.0 (10.04) | 5753 | 37.64 | 15283 | 2.10E-04 |